\documentclass[fleqn,10pt]{wlscirep}
\usepackage[utf8]{inputenc}
\usepackage[T1]{fontenc}

\usepackage{amsmath}
\usepackage{subcaption}
\usepackage{float}

\DeclareMathOperator*{\argmax}{argmax}
\title{Social Network Structure is Predictive of Health and Wellness}

\author[1,2]{Suwen Lin}
\author[1,2]{Louis Faust}
\author[1,2]{Pablo Robles-Granda}
\author[1,2,*]{Nitesh V. Chawla}
\affil[1]{Department of Computer Science and Engineering, University of Notre Dame, Notre
Dame, IN 46556, USA}
\affil[2]{Interdisciplinary Center for Network Science and Applications,
University of Notre Dame, Notre Dame, IN 46556, USA}

\affil[*]{Corresponding author. Email: nchawla@nd.edu}

\begin{abstract}

Social networks influence health-related behaviors, such as obesity and smoking. While researchers have studied social networks as a driver for diffusion of influences and behaviors, it is less understood how the structure or topology of the network, in itself, impacts an individual's health behaviors and wellness state. In this paper, we investigate whether the structure or topology of a social network offers additional insight and predictability on an individual's health and wellness.  We develop a model called the Network-Driven health predictor (NetCARE) that leverages features representative of social network structure. Using a large longitudinal data set of students enrolled in the NetHealth study at the University of Notre Dame, we show that the NetCARE model improves the overall prediction performance over the baseline models -- that use demographics and physical attributes -- by 38\%, 65\%, 55\%, and 54\% for the wellness states -- stress, happiness, positive attitude, and self-assessed health -- considered in this paper. 
\end{abstract}
\begin{document}

\flushbottom
\maketitle
%
%
\thispagestyle{empty}


\section*{Introduction}

Social networks play an important role in the diffusion of behavior, attitudes, tastes, and beliefs. Several studies have shown that such characteristics leverage the existing social connections and ties for diffusion. This phenomenon is demonstrative of the similarity or \emph{homophily} between the nodes in the network (ego and alter, for example) but also of the social influences that affect people. Some examples of this diffusion process include: the spread mechanism of diverse health conditions over social networks -- such as obesity \cite{christakis2007spread} and smoking \cite{christakis2008collective}, the effect of social network on personal psychological traits -- such as affection \cite{bearman2004chains} and happiness \cite{fowler2008dynamic}, and the spread of health behaviors through social networks \cite{centola2010spread}.
People's interactions through social networks or social media platforms have also been used to discover aspects of emotions experienced by individuals \cite{stieglitz2013emotions}, mental illness \cite{coppersmith2014quantifying,reece2017forecasting}, and activity patterns \cite{weerkamp2012activity}. Different social network types, such as friendship or non-friendship networks, can also provide insights to predict mental health in adults \cite{fiori2006social}.


While these studies have shown the prevalence of social network effects in diffusion of behaviors, attitudes, and beliefs, there are also other self-selection and external factors that influence the similarities in behaviors and attitudes (context, culture, dispositions). Consider the network-effect hypothesis. It suggests that similarities in lifestyle and health practice, including health behavior, moods, emotions, cultural norms, etc. \cite{cohen1985social, cohen2008detecting}  among individuals is also a result of influence and diffusion within their network through their ties.  In addition, the self-selection hypothesis suggests that ties among people are driven by similar pre-dispositions to attitudes or beliefs or behavior, so those factors might even be driving the formation of the tie \cite{mcpherson2001birds,newman2003mixing}. But how do the network and self-selection hypotheses help explain one's health and wellness state? We hypothesize that both are critical to explaining and predicting an individual's wellness state. There are self-selection and external factors (such as demographics and an individual's propensity for being active) as well as network factors (ties derived through communication networks) that collectively lead to an improved wellness state of an individual.  We break down this hypothesis into two research questions in this paper that also drive the core contributions.

\textbf{RQ1: Is social network structure indicative of health behavior? (Analysis)}

\begin{figure}[H]
\centering
\begin{subfigure}{.5\textwidth}
  \centering
  \includegraphics[width=\linewidth]{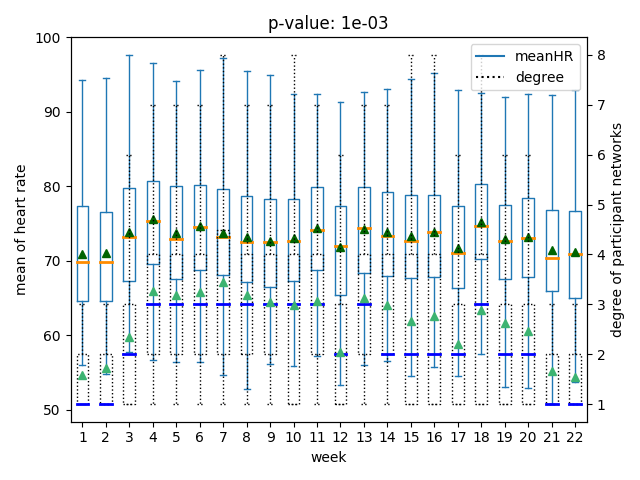}
  \caption{The relation between network structure and health behavior}
  \label{showcase1}
\end{subfigure}%
\begin{subfigure}{.5\textwidth}
  \centering
  \includegraphics[width=\linewidth]{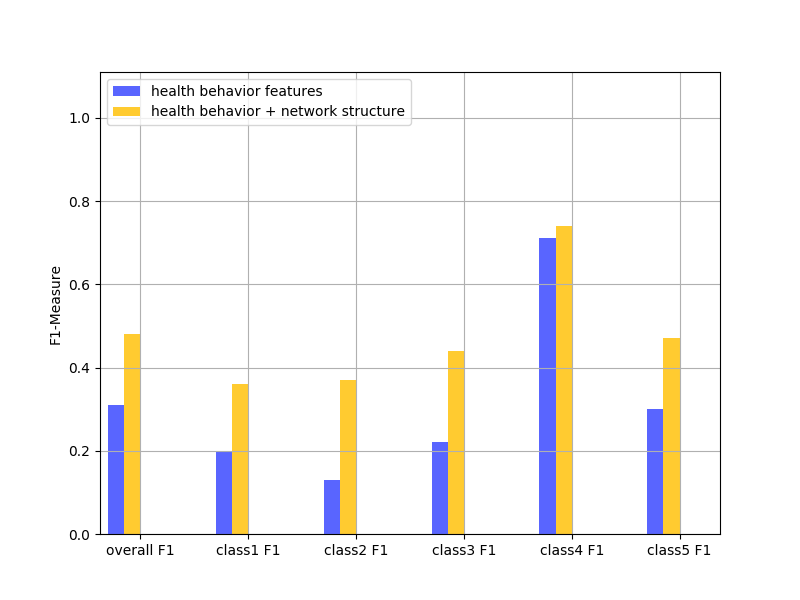}
  \caption{Positive attitude prediction performance including overall F1-Measure and within-class F1-Measure}
  \label{showcase2}
\end{subfigure}%
\caption{Main results for the research questions}
\label{showcase}
\end{figure}

We consider social network structure to be represented by social network properties such as  node degree, clustering coefficients, and centrality. We consider the health and wellness behavior as data captured from wearable devices --- heart rate, daily steps, and activity states --- and survey data. We analyze, quantitatively and qualitatively, the relationship between the social network structure and the aforementioned health information.  An example of this relation is shown in Figure \ref{showcase1}. This figure shows how the node degree on the network (shown in dashed box-plot) is related to the changes of the heart rate (shown with regular-lines). The figure shows that these values seem positively correlated because, as time progresses, the mean and the median of node degree (shown as blue lines and as green triangles, respectively) increase or decrease when the mean or median of the heart rate (show as orange lines and dark green triangles, respectively) also increase or decrease, where the corresponding normalized cross correlation is 0.84. In this paper we will provide evidence that social network structure contains information that captures the change in statistics of health behaviors. 


\textbf{RQ2: How predictable are the wellness states from the incorporation of social network structure? (Prediction)}

While previous research has shown that health behavior data captured from wearables is indicative of diseases or symptoms of diseases  \cite{muaremi2013towards,stutz2015smartphone}, in this paper we incorporate the social network structural features in addition to health behavior data captured by wearables using a machine learning framework that predicts aspects of health and wellness (NetCARE). 
  We consider various health and wellness states such as stress, happiness, positive attitude, and self-assessed health indicators. 
  Figure \ref{showcase2} summarizes the improvement of overall F1-Measure and within-class F1-Measure for positive attitude prediction by involving the network structural information. Clearly, the knowledge of social network structure provides significant improvement over using the data from the wearables and / or the individual's demographics alone. 

\section*{Methods}

\subsection*{Data}

\begin{figure}[ht]
\centering
\includegraphics[width=\linewidth]{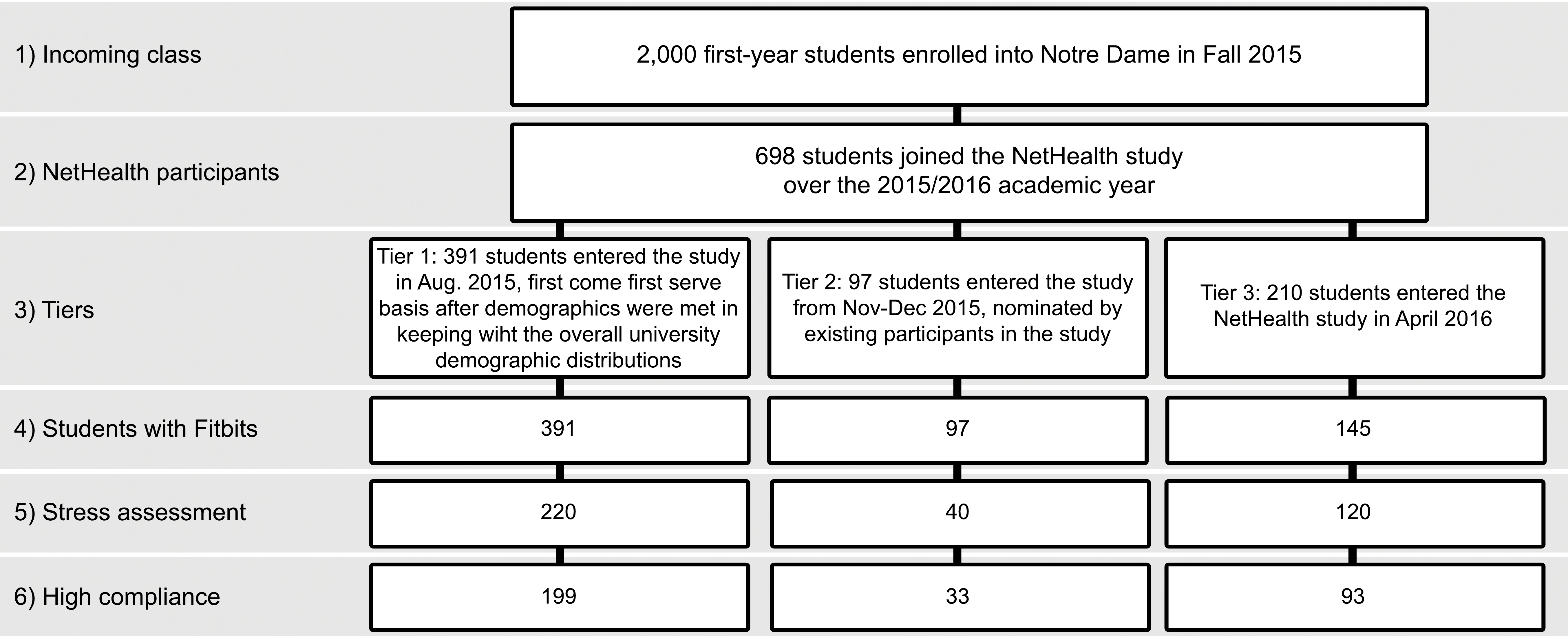}
\caption{Consort diagram of NetHealth recruitment and students selected for this analysis in this paper.}
\label{consort}
\end{figure}

We use data from the NetHealth study \cite{netHealthWebsite}, an ongoing project at the University of Notre Dame, collecting survey, phone and Fitbit data from an initial cohort of 698 first-year students who enrolled in the Fall of 2015 \cite{purta2016experiences}. 
All procedures were fully approved by the University of Notre Dame Institutional Review Board before distribution and performed in accordance with the relevant guidelines and regulations. All study participants provided informed consent and acknowledged all of the study goals, procedures, and data privacy, prior to any data collection. 
An outline of the recruitment process and student sample numbers is provided in Figure \ref{consort}. Participants were provided with a Fitbit Charge HR and had an app installed on their phone, which was leveraged to build the social network on the basis of communication patterns (call, message). Each participant was anonymized and given a unique hashID. For the purpose of the study, the data is organized into three parts: survey data, social network data, and Fitbit data.

Students were required to complete an entrance survey before arriving on campus and follow-up surveys after each semester. Surveys contain a battery of questions regarding individual demographics and self-reported mental and physical wellness assessments. 
It should be noted that the survey questions are different for each semester and not all students took part in all the surveys. 
We consider the following three datasets from NetHealth:

\emph{Social Network Data.}. The students' social networks were built using their communication activities including texts and phone calls captured through an app. The app can automatically gather the time, source and destination of their communication activities. As for phone calls, the app can also record the duration of the call and whether the call was answered.  

\emph{Health Behavior and Demographic Data.} This physical data is obtained from the Fitbit metrics which include health-related behavioral variables such as heart rate, step and activity states. 
Besides the minute-by-minute raw heart rate and step data, Fitbit also separates and tracks four heart-rate zones (out of range, fat burn, cardio, and peak) \cite{fitbit2016heart}. As for the activity records, Fitbit calculates activity states per minute based on METS, a weight-agnostic measure of activity, that represent sedentary, lightly active, fairly active, and very active activity states \cite{fitbit2016active}. We also consider the gender of the participants in our analysis (the only demographic feature).

\emph{Wellness State.}
These data are from surveys answered by participants each academic semester. Due to the different survey questions across semesters, we cannot jointly analyze all the surveys. For that reason, we selected the survey taken in Fall 2016, which contains questions about wellness states - stress, happiness, positive attitude and self-assessed health - and covers most of our participants (380 subjects) as this is the first semester in which all three tiers are present in the study. Accordingly, we considered contemporary data from fitness trackers and social interaction from August 2016 to December 2016. We excluded 47 participants for missing Fitbit or social network data. As a result, our data covers 325 participants.
Table \ref{baseline} shows how these data samples are chosen in the NetHealth study from the view of their demographics, where some of the missing data correspond to race and age features. Table \ref{survey} presents the questions from the survey, their possible answers and corresponding compositions of the four areas of interest of our study. Stress is categorized into four categories from 1 to 4, with category 1 indicating lowest level stress and category 4 indicating the highest level of stress. Likewise, happiness and self-health report are also divided into four categories, with a numeric increase in the level from category 1 to category 4. Finally, given the distribution of positive attitude, 
we split it into five categories, with category 1 indicating the least positive and category 5 indicating the most. 

\begin{table*}[]
\centering
\begin{tabular}{llll}
\multicolumn{2}{l}{\textbf{demographic}} & \textbf{\# Data Points}& \\
\textbf{gender} & male &  146 (45\%) &  \\
\textbf{} & female & 179 (55\%) & \\ \cline{1-3}
\textbf{race} & white & 227 (70\%) &  \\
\textbf{} & latino & 36 (11\%) & \\
\textbf{} & asian & 29 (9\%) & \\
\textbf{} & black & 18 (6\%) & \\
\textbf{} & foreign & 14 (4\%) & \\
\cline{1-3}
\textbf{age} & 17 & 36 (11\%) &  \\
\textbf{} & 18 & 182 (56\%) & \\
\textbf{} & 19 & 11 (3\%) & \\
\cline{1-3}

\end{tabular}
\caption{\label{baseline} Summary of demographics in data samples. The total number of samples are 325. }

\end{table*}

\begin{table*}[ht]
\centering
\begin{tabular}{llcc}
\multicolumn{2}{l}{\textbf{Questions}} & \textbf{Level} & \textbf{\#Participants (\# Male, \# Female)} \\ \cline{1-4}
\textbf{Stress} & Felt nervous and stressed &  1 &  14 (12,2)  \\
\textbf{} &  & 2 & 95 (58,37) \\
\textbf{} &  & 3 & 134 (52,82) \\
\textbf{} &  & 4 & 82 (24,58) \\ \cline{1-4}
\textbf{Happiness} & Would you say that you are very happy &  1 & 40 (14,26)  \\
\textbf{} &  & 2 & 84 (38,46) \\
\textbf{} &  & 3 & 144 (64,80) \\
\textbf{} &  & 4 & 57 (30,27) \\ \cline{1-4}
\textbf{Positive attitude} & I take a positive attitude toward myself &  1 & 2 (0,2) \\
\textbf{} &  & 2 & 23 (8,15) \\
\textbf{} &  & 3 & 65 (28,37)  \\
\textbf{} &  & 4 & 160 (72,88) \\
\textbf{} &  & 5  & 75 (38,37) \\ \cline{1-4}
\textbf{Health} & Health Rating &  1 & 7 (4,3)  \\
\textbf{} &  & 2 & 56 (18,38) \\
\textbf{} &  & 3 & 200 (95,105) \\
\textbf{} &  & 4 & 62 (29,33) \\ \cline{1-4}
\end{tabular}
\caption{\label{survey} Summary of wellness-related survey questions in NetHealth study. The total selected participants are 325.}
\end{table*}

\subsection*{Data Preprocessing}
There are two steps for data preprocessing. First, to ensure there is no bias between the students with different 
levels of the sparsity of daily Fitbit data, 
we eliminated samples with less than 80\% daily compliance in our analysis because 80\% daily compliance provides reasonable estimates of students activity \cite{purta2016experiences}. Second, we divided the data from August 2016 to December 2016 using a one-week window, and each data point includes one-week of Fitbit data, social data and the corresponding survey data from the Fall 2016 survey. The reason for this data partition is that the survey is taken in Fall 2016, but health behavior and social network data span five months from August to December 2016. Subsequently, one has to then select the time range for corresponding data streams from the Fitbit or the social network.  
Too fine-grained an interval over time like a day or an hour is not effective for the measurement of the social network since the daily or hourly social interactions for one person are limited. Too coarse-grained an interval also has negative effects due to possible information loss if an undirected and unweighted graph is used to represent the network.
%
%



\subsection*{Feature Extraction}
We extracted several features from the data streams to build an appropriate feature vector for the learning algorithms as we describe next. We also included gender information in the feature vector, as we wanted to study the influence of gender in predictability of the health behaviors.


\subsubsection*{Gender Information}
The World Health Organization recognized there are gender differences in stress-related syndromes \cite{world2002gender}. For example, women have much higher incidence rates of stress than males. Based on this insight, we extracted the gender information from the survey data to use as an additional feature in the feature vector. Table \ref{survey} shows the population distributions for the different levels of survey variables for males and females. Specially, consider the case of stress, happiness, and health, males mainly fall into level 2 and 3 and most of the females fall into level 3. We use gender as an independent variable in our analysis (predictor). 



\subsubsection*{Health Behavioral Data}
We categorize the physical attributes captured from Fitbit (heart rate, steps, and activity states) as health behavior data. This data is segmented into the weekly intervals discussed in the previous section. Then, summary statistics of mean and variance (or standard deviation) are computed on these temporal segments. 

\textbf{Heart Rate}.
We computed the mean and variance for the heart rate over each week for each of the participants. We also applied ANOVA tests to examine the heart rate differences among different stress levels, happiness levels, health levels, and positive attitude levels. 
Results showed significant differences of heart rates for different stress levels ($p < .001$), happiness levels ($p < .001$), health levels ($p < .001$), and positive attitude levels ($p < .001$).




\textbf{Steps}.
The raw data for steps are also recorded minute by minute, but it is more likely to be zero for most of the minutes in one week due to the nature of walking. Thus, we first transformed the raw step data into the sum of steps each day. Then, we computed the mean and standard deviation of these daily steps for each week and each person as features.

\textbf{Activity State}.
Fitbit tracks the users' activities and records their corresponding pre-defined states every minute. There are 4 possible states: sedentary, lightly active, fairly active and very active. So, the sum of minutes in each state on each day are computed first, then mean and standard deviation of these daily summations for each state within each week are computed. 

\subsubsection*{Social Network Data}

Social networks were constructed from the communication patterns of phone calls and text messages. To avoid spurious connections (such as spam), we eliminated communication edges that had a frequency of less than three times within a 5-month period. The NetHealth study collected communication data not only from within all the participants but also between participants and people from outside of the study as well. As a result, we had two types of social networks: one that includes all the data (\textit{whole network}) the one that only includes communication data of the participants within the study (\textit{participant network}). 
The \textit{participant network} only includes friends or classmates since all the participants are undergraduate students with the same class-standing or year in the same university.  We regard the \textit{participant network} as a \textit{friend network}, which is one of the five types of social networks that can affect health \cite{fiori2006social}. However, we studied both the whole and participant networks in our analysis.


As we mentioned before, each time step in our social network analysis consists of one week. The social networks are undirected and unweighted representations of communication patterns for each week.  We then derive several features that are representative of the social network structures, including  network degree \cite{diestel2005graph}, number of triangles, clustering coefficient \cite{watts1998collective}, betweenness centrality \cite{freeman1977set}, and closeness centrality \cite{sabidussi1966centrality} for each person in the network.

\subsection*{Analysis framework}
\subsubsection*{Health Behavior Relationship Analysis}
To answer RQ1, we investigate if there is a relationship between social network structure and health behavior and whether the social network structure properties are predictive of the health behaviors. Specifically,  we examine the relationship between social network topological properties including degree, number of triangles, clustering coefficient, closeness centrality and betweenness centrality for each node (individual) in the \textit{participant network} and in the \textit{whole network} and health behavioral variables including heart rates, steps and activity states.

We visualize all the 22 weeks using box plots to show the relationship between network structural variables and health behavioral statistics in a qualitative way. We use cross correlation coefficients \cite{yoo2009fast} to capture the correlation between each of the behavioral variables and each of the network structural variables. It should be noted that the links from physical and behavioral variables to social network variables vary across individuals. Thus, in this study we compute the correlation for each individual, and then we sum up the total number of coefficients showing a value greater than 0.5. 


\subsubsection*{Wellness State Prediction}

\begin{figure}[ht]
\centering
\includegraphics[width=0.6\linewidth]{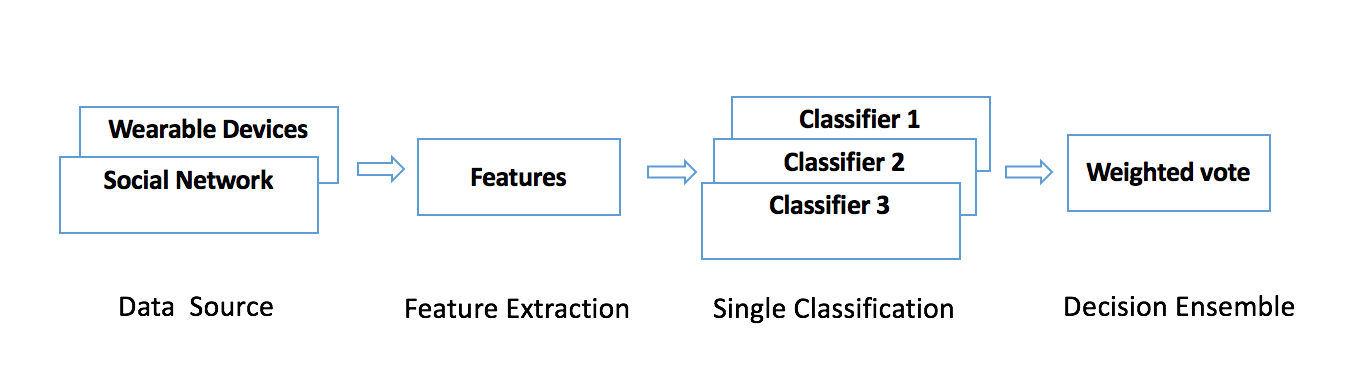}
\caption{A network-driven prediction architecture, NetCARE.}
\label{arch}
\end{figure}

To answer RQ2 we propose NetCARE, a network-driven prediction architecture, to make full use of social network structure features in health prediction problems. Figure \ref{arch} shows the schema of NetCARE. The algorithm incorporates social network structure, wearable data, and demographic data as independent variables of a machine learning model. This algorithmic architecture allows us to select network features and add other data sources as needed. It also ensures the flexibility to modify, extend, or add classifiers. 

Let us use stress detection as an example to explain this architecture in detail.
We formulate the problem as one of 4-categories classification, where features  are extracted from health behaviors and network structure, and the class categories are the levels of stress. We then employed five popular classifiers for the problem: K-Nearest Neighbors (KNN), Classification and Regression Trees (CART), Support Vector Machines (SVM), Logistic Regression (LR) and Random Forests (RF). The data was divided into 75\% for training and 25\% for testing. We used 5-fold cross-validation to find the hyper-parameters of the algorithms, and used grid search to find the combination of those parameters that achieved the best performance. We then consider the  averaged F1-Measures for all levels and those within each level as a metric of performance. 
Specifically, we tuned the number of neighbors for KNN and the leaf size for CART. For SVM, we conducted experiments over the three different kernels: polynomial, linear and radial basis function (RBF) kernels, where various degrees of polynomial kernels were also taken into consideration. For LR, we searched on different values for regularization coefficients and the learning rate of the optimization algorithm. For RF, we conducted experiments with different numbers of trees from 10 to 100 at increments of 5. We use 35 trees for the results reported in this document. 

Furthermore, we applied an ensemble method with a weighted voting \cite{dietterich2000ensemble} scheme to improve the performance of the individual classifiers. We chose three single classifiers as base classifiers. 
We used the notation $p_{ij}$ to represent the probability where classifier $i$ classifies the input instance $x$ as class $j$. So, given the weight $w_{ij}$ assigned to probability $p_{ij}$, the ensemble rule to get the output $y_{vote}$ can be formulated as (\ref{vote}). For weight selections, we used cross-validation over training data to select the best weights $\{w_{ij}^{optimal}\}$ from all possible combination of weight $w_{ij}$ from 0 to 1 with interval 0.1. 
Thus the final decision was made by the label that maximizes the averall weighted vote as defined by eq. (\ref{rule}).
%
%
\begin{equation}\label{vote}
\begin{split}
    y_{vote}(\{w_{ij}\}) = \argmax_j{\sum_{i=1}^{3}w_{ij}p_{ij}} 
    \qquad s.t. \quad \sum_{j=1}^4{w_{ij}}=1,\text{ } j \in \{1,2,3,4\}, \text{ and  } i \in \{1,2,3\}
\end{split}
\end{equation}
%
%
\begin{equation}\label{rule}
    y_{final} = \argmax_j{\sum_{i=1}^{3}w_{ij}^{optimal}p_{ij}} \qquad s.t. \text{ } j \in \{1,2,3,4\}, \text{ and } i \in \{1,2,3\}
\end{equation}
\section*{Results}

In order to test our framework, we performed two sets of experiments. First, we evaluated the interactions among the variables associated with social network structures and those related to health behavior. The objective of this analysis was to validate whether those interactions were meaningful. Second, we used our framework to predict various wellness variables. We compared the performance of our framework with a baseline that considers either health-behavior data and social network features in isolation. We did this to verify our hypothesis that combining network effects and self-similarity will lead to better predictions.

\subsection*{Relationship Analysis Between Social Network Structure and Health Behavior }

\begin{figure}[H]
\centering
\begin{subfigure}{.5\textwidth}
  \centering
  \includegraphics[width=\linewidth]{meanHR_degree.png}
  \caption{box plot of mean of heart rate and degree of \textit{participant network}}
  \label{fig:core1}
\end{subfigure}%
\begin{subfigure}{.5\textwidth}
  \centering
  \includegraphics[width=\linewidth]{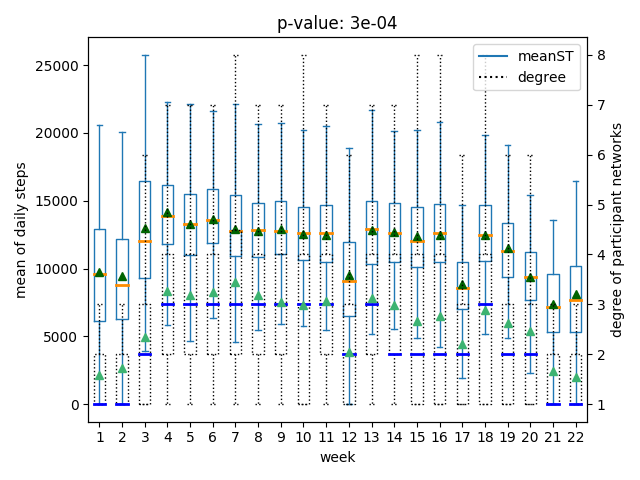}
  \caption{box plot of mean of daily steps and degree of \textit{participant network}}
  \label{fig:core2}
\end{subfigure}
\begin{subfigure}{.5\textwidth}
  \centering
  \includegraphics[width=\linewidth]{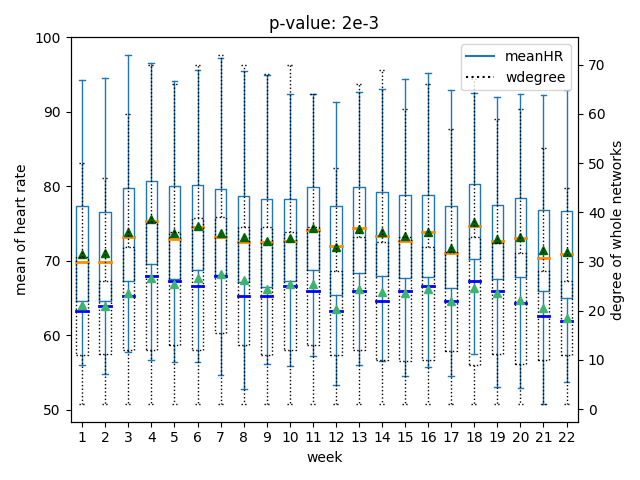}
  \caption{box plot of mean of heart rate and degree of \textit{whole network}}
  \label{fig:whole1}
\end{subfigure}%
\begin{subfigure}{.5\textwidth}
  \centering
  \includegraphics[width=\linewidth]{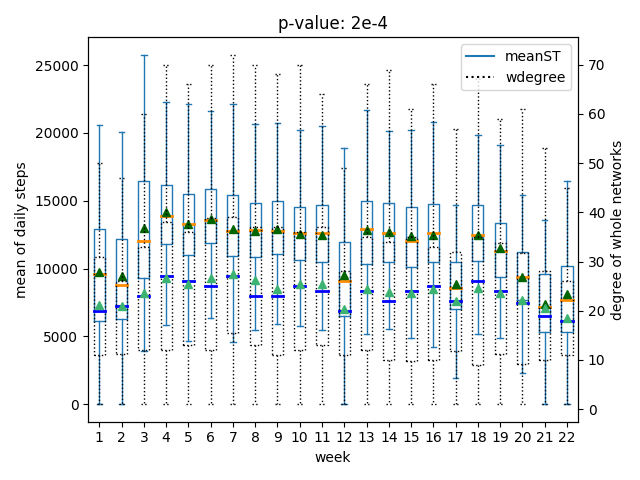}
  \caption{box plot of mean of daily steps and degree of \textit{whole network}}
  \label{fig:whole2}
\end{subfigure}
\caption{Relation between social network structures and health behaviors}
\label{fig:core}
\end{figure}

We analyzed the relations among the physical and social network variables. To this goal, we visualized all the possible pairs of 5 structural features of social networks in 2 networks (\textit{whole network} and \textit{participant network}) and 6 physical and behavioral features with box plots, which resulted in 60 box plots. For each box in a plot, we present the distribution of a physical-behavioral feature and a social network structure feature for all the participants over 22 weeks. Note that health-behavioral features for each week were extracted as mean values from the raw data. For example, Figure \ref{fig:core1} presents the distribution of average heart rates for all participants and the node degree distribution over 22 weeks of the \textit{participant network}. Figure \ref{fig:core1} and \ref{fig:core2} represents the relationships 
for the \textit{participant network} structures, while the relationships for the \textit{whole network} are presented in Figure \ref{fig:whole1} and \ref{fig:whole2}. The remaining box plots can be found in supplemental materials. 


As shown in Figure \ref{fig:core}, the median and the mean of health behavior data for each week (dark orange lines and dark green triangles in the figures, respectively) change over time and the median and mean of network properties (dark blue lines and sea green triangles in the figures, respectively) follow a similar pattern over time.

We performed tests to verify whether there was a statistically significant difference between the distribution of health behavior features and the network structure features across high and low-value ranges. Specifically, using t-tests we tested for whether the feature values representative of behavioral data varied in the strength of the relationship with the network data ranges. For example, consider the relationship between daily steps and network degree. The derived p-value of 0.0003 shows that there is a significant difference between the daily steps in conjunction with higher network degree versus daily steps in conjunction with lower network degree. 
Figures~\ref{fig:core2} show the results. 
After correcting for multiple tests \cite{dunn1961multiple}, our results show that social network properties have significant relationships 43 out of 60 times, supporting the hypothesis that social networks are indicative of changes in health behavior. 





Further, we used cross correlation coefficients \cite{yoo2009fast} to quantify the extents to which the network structure features can reflect the information flow of health behaviors. After calculating the means of each feature from health behaviors and network structure for every week, we computed the coefficients of the means of health behavior features and the means of structure features (60 pairs). The results showed 43 of 60 pairs with a absolute correlation coefficient that is no less than 0.5 and 28 of the 60 pairs with absolute correlation no less than 0.7. Table \ref{spear} shows the results from all the pairs of behavioral features and network structure features. 

The correlation from network-structure features to the very active state is generally stronger by about 0.3 on average than that from structure to sedentary, fairly active or lightly active states. 
%
We noticed none of the structure features have a strong relation to the data of the lightly active state. After excluding results of the lightly active state, we found that the number of triangles in \textit{participant network} and \textit{whole network} can be a good indicator of the change of other health behavior data. Also, except lightly active, each behavioral feature could be related to at least one of the structural features with absolute coefficients no lower than 0.7. Especially, the correlation coefficient between degree in the \textit{whole network} and steps is almost 0.9. 
We did the same experiments on the medians of each feature for every week. We found 37 of 60 pairs had correlations no less than 0.5. 
Comparing the results from metrics in \textit{participant network} with those from metrics in \textit{whole network} in the table, we could not identify major differences between the two networks. For example, the coefficient between the fairly active state data and Closeness Centrality in the \textit{participant network} is almost half of coefficient between fairly active data and Closeness Centrality in the \textit{whole network}, while the coefficient between fairly active data and Clustering Coefficients in the \textit{participant network} almost doubles the coefficient between fairly active data and Clustering Coefficients in the \textit{whole network}. This finding suggests that there are different effects in the two network types and it is necessary to include both networks in our analysis.
In summary, table \ref{spear} show that the network structure seems to capture the changes of health behavior -- although in lesser extent with respect to the lightly active state. 

\begin{table}[!htp]
\centering
\begin{tabular}{|c|c|c|c|c|c|c|}
\hline
&heart rate & steps & sedentary & lightly active &fairly active &very active \\
\hline
Degree in \textit{participant network}&0.84&0.89&-0.44&-0.014&0.49&0.87\\
Number of triangles in \textit{participant network}&0.74&0.83&-0.61&0.24&0.68&0.79\\
Clustering Coefficient in \textit{participant network}&0.65&0.75&-0.51&0.15&0.59&0.66\\
Betweenness Centrality in \textit{participant network}&0.78&0.68&-0.19&-0.20&0.20&0.72\\
Closeness Centrality in \textit{participant network}&0.83&0.85&-0.32&-0.14&0.35&0.86\\\hline
Degree in \textit{whole network}&0.81&0.90&-0.57&0.15&0.62&0.88\\
Number of triangles in \textit{whole network}&0.79&0.89&-0.62&0.23&0.69&0.85\\
Clustering Coefficient in \textit{whole network}&0.83&0.79&-0.32&-0.12&0.35&0.79\\
Betweenness Centrality in \textit{whole network}&-0.76&-0.85&0.65&-0.28&-0.71&-0.79\\
Closeness Centrality in \textit{whole network}&0.75&0.78&-0.59&0.24&0.65&0.71\\

\hline
\end{tabular}
\caption{\label{spear}Normalized cross correlation coefficients of each pair of health behavior feature averages and social network structure feature averages.}
\end{table}

We also evaluated variable interactions for each participant in the dataset.
This was done to evaluate changes in health behavior per individual.
Each analyzed sample point is the average of the behavior over one week per person. 
We counted the total number of persons with more than 0.5 on absolute cross correlation coefficients to show the extent to which the structural features can capture the changes of health behavior. 
Table \ref{sep} lists the numbers of participants with medium to strong correlation for each pair of health behavioral features and network features. In the table, each health behavior feature can be related to one of structural features for both \textit{whole network} and \textit{participant network} for at least 20\% of participants. The table shows that Closeness Centrality in either the \textit{whole network} or \textit{participant network} capture a relationship with steps for over 42\% of the samples. These results imply an underlying relationship between social network structures and health behaviors.

Table \ref{corre} summarizes the fraction of participants with medium to strong correlations with respect to health behavior features and at least one graph structure feature
from \textit{participant network}, \textit{whole network} or both networks, respectively. 
Particularly, the percentages are the fraction of persons whose health behavior data has no less than 0.5 cross correlation coefficients with any of the network structure features.
For example, the person whose any one of 5 features from \textit{participant network} is related to steps data with coefficients no less than 0.5, is counted. 
According to the first and second columns in Table \ref{corre}, the three metrics from each kind of social networks can closely capture the changes of some health behaviors for about 50\% of the participants. Specifically, 74\% of the participants have higher correlations between steps and one aspect of structural features. The last column in Table \ref{corre} shows that both types of social networks maintain information of the time-varying heart rate averages, step averages, and averaged minutes in each activity states, for over 50\% participants. Additionally, the fraction of persons with correlation coefficients no less than 0.7 between the number of steps and the structural features in \textit{whole network} or in \textit{participant network} is 40\%, and between the mean of heart rates and the structural features is 26\%. These results imply the \textit{whole network} contains more sufficient information about health behaviors than the \textit{participant network}, but both of them are essential pieces, given the increase of numbers in the last column.

\begin{table}[ht]
\centering
\begin{tabular}{|c|c|c|c|c|c|c|}
\hline
&heart rate & steps & sedentary & lightly active &fairly active &very active \\
\hline
Degree in \textit{participant network}&52&86&43&39&42&47\\
Number of triangles in \textit{participant network}&31&38&29&24&37&26\\
Clustering Coefficient in \textit{participant network}&28&37&27&24&35&27\\
Betweenness Centrality in \textit{participant network}&38&38&34&35&32&34\\
Closeness Centrality in \textit{participant network}&99&145&81&79&82&98\\\hline
Degree in \textit{whole network}&78&94&81&58&70&68\\
Number of triangles in \textit{whole network}&63&100&69&43&62&60\\
Clustering Coefficient in \textit{whole network}&47&66&46&37&51&39\\
Betweenness Centrality in \textit{whole network}&50&52&57&54&53&49\\
Closeness Centrality in \textit{whole network}&91&137&98&83&94&80\\
\hline
\end{tabular}
\caption{\label{sep}Number of persons whose health behaviors have medium to strong correlation with social network structures.}
\end{table}

\begin{table}[ht]
\centering
\begin{tabular}{|c|c|c|c|}
\hline
health-related data &\textit{participant network} (\%)&\textit{whole network} (\%)&both network (\%)\\
\hline
heart rate & 133 (41)& 157 (48)& \textbf{193 (59)}\\
\hline
steps & \textbf{182 (56)}& \textbf{204 (63)}& \textbf{239 (74)}\\
\hline
sedentary & 125 (38)& \textbf{166 (51)}& \textbf{202 (62)}\\
\hline
lightly active  & 132 (41)& 145 (45)& \textbf{199 (61)}\\
\hline
fairly active  & 122 (38)& \textbf{164 (50)}& \textbf{194 (60)}\\
\hline
very active  & 133 (41)& 143 (44)& \textbf{186 (57)}\\
\hline
\end{tabular}
\caption{\label{corre}Summary of subjects with medium to strong correlation to the social network structures. Percentages are the fraction of persons whose health behavior data has no less than 0.5 cross correlation coefficients with any of the network structure features, where total number of persons in the data is 325.}
\end{table}



In summary, these experiments verify the interactions among network-structure variables and health-behavioral variables. Specifically: \textbf{1)} We visualized the effect of the network structures on the health behaviors and discovered the network structures can qualitatively capture the changes of behavioral variables. \textbf{2)} We conducted t-tests to check if higher values of health-behavior variables corresponded higher values of structural variables and are different from lower values of both types of variables. Our results showed 43 out of 60 with significant differences after Bonferroni corrections. \textbf{3)} We used normalized cross correlation coefficients to describe the role of network structures in statistics. The results showed 43 of 60 pairs of behavior features and structural features with a correlation coefficient that is no less than 0.5 and about half of the pairs with a coefficient no less than 0.7. \textbf{4)} We analyzed the variable interactions of structure and physical features at the individual level. We found that up to 145 out of 325 participants showed a high correlation between their Closeness Centrality of networks and steps, and up to 74\% of the participants showed the similar relation between the aggregated network features and steps.

\subsection*{Predicting Wellness State}

After implementing the five single classifiers, we chose SVM, KNN, and RF, to create our ensemble learning model. The first five rows of Table \ref{health} show the performance of our ensemble classifiers for stress prediction. We report the F-score for all stress levels and each level. The table shows that social network variables alone are comparable or even a little better to health behavior data for overall F-score and stress level 3, while others are worse. Thus, we suspect social network structures contain information about stress from a complementary perspective compared to that of health-behavior variables, i.e. there seems to exist an underlying relationship between social network structures and stress states.  
The table also shows that joining features from social networks and health behaviors improve predictions as evaluated by the F1-score improvement on both the combined performance and the individual performance per stress level. 
The most noticeable improvement corresponds to stress level 1.  
%

Additionally, we perform the same analysis for other health and wellness variables. In particular, we assess the effect of combining social network structure variables and health-behavior variables to predict happiness, positive attitude and self-assessed health (Table \ref{survey}). These results are also shown in Table \ref{health}. As in the case of stress, the table shows that using social network structures can improve prediction performance for these 3 health and wellness variables. Table \ref{health}, shows that our NetCARE framework provides improvements of: 1) 65\% and up to 617\% on the overall F1-Measure and the within class F1-Measure of \emph{happiness}, respectively; 2) 55\% and up to 185\% on the overall F1-Measure and the within class F1-Measure of \emph{positive attitude}, respectively; and 3) 54\% and up to 200\% on the overall F1-Measure and the within class F1-Measure of \emph{self-assessed health}, respectively.
These results provide evidence that not only that structural features could be helpful in applications of wellness state prediction and health perceptions, they also show a potential relationship between wellness variables and social network structure.






 

\begin{table}[!ht]
\centering
\begin{tabular}{|c|c|c|c|c|c|c|c|}
\hline
\textbf{Stress Prediction} &  F1 & Level1 & Level2 & Level3 & Level4 &\\
\hline
gender + health behavior data & 0.42 & 0.18 & 0.53 & 0.64 & 0.34&\\
\hline
social network structures & 0.34 & 0.05 & 0.43 & \textbf{0.63} & 0.26&\\
\hline
gender + health behavior data + social network & 0.58 & 0.46 & 0.63 & 0.70 & 0.55 &\\
\hline
improvement  & 38\% & \textbf{155\%} & 19\% & 9\% & 62\% &\\
\hline\hline

\textbf{Happiness Prediction} &  F1 & Level1 & Level2 & Level3 & Level4 &\\
\hline
gender + health behavior data & 0.31 & 0.06 & 0.31 & 0.62 & 0.24 &\\
\hline
social network structures & 0.21 & 0.00 & 0.2 & 0.60 & 0.02 &\\
\hline
gender + health behavior data + social network & 0.51 & 0.43 & 0.52 & 0.67 & 0.44 &\\
\hline
improvement & 65\% & \textbf{617\%} & 68\% & 8\% & 83\% &\\
\hline\hline

\textbf{Positive Attitude Prediction}& F1 & Level1 & Level2 & Level3 & Level4 & Level5\\
\hline
gender + health behavior data & 0.31 & 0.20 & 0.13 & 0.22 & 0.71 & 0.30 \\
\hline
social network structures & \textbf{0.40} & \textbf{0.70} & 0.08 & 0.23 & 0.70 & 0.25 \\
\hline
gender + health behavior data + social network &0.48 & 0.36 & 0.37 & 0.44 & 0.74 & 0.47 \\
\hline
improvement & 55\% & 80\% & \textbf{185\%} & \textbf{100\%} & 4\% & 57\% \\
\hline\hline

\textbf{Self-assessed Health Prediction} & F1 & Level1 & Level2 & Level3 & Level4 &\\
\hline
gender + health behavior data   & 0.35 & 0.29 & 0.13 & 0.77 & 0.20 &\\
\hline
social network structures   & 0.21 & 0.00 & 0.05 & 0.77 & 0.00 &\\
\hline
gender + health behavior data + social network   & 0.54 & 0.6 & 0.39 & 0.79 & 0.37 &\\
\hline
improvement  & 54\% & \textbf{107\%} & \textbf{200\%} & 3\% & 85\% &\\
\hline
\end{tabular}
\caption{\label{health}Prediction results for happiness, positive attitude and self-assessed health}
\end{table}

\section*{Discussion}
The main contributions of this paper can summarized as follows:
\begin{itemize}
    \item We discovered that social network structure is correlated with health behavior data captured from wearables, and can capture the trends. This relationship between social network structure and health behavior is statistically significant.  
    \item We demonstrated that social network structure is essential to improved predictability on wellness states. This result is of importance as just relying on data derived from wearables and demographics does not express a complete picture about an individual, and one's social network is an essential element to understanding and predicting health and wellness. 
\end{itemize}

%
%

Social network analysis has been used for health-related problems including mental health \cite{stieglitz2013emotions, fowler2008dynamic}, physical well-beings \cite{christakis2007spread, christakis2008collective}, and illness \cite{reece2017forecasting, wenger1997social}. 
Most of the work has largely focused on social networks as a diffusion mechanism of health \cite{christakis2007spread,christakis2008collective,bearman2004chains,fowler2008dynamic,centola2010spread} or emotions \cite{stieglitz2013emotions,coppersmith2014quantifying,reece2017forecasting,weerkamp2012activity}. This paper provides a novel perspective on the value of social network structure in not only understanding our health behavior but also in predicting the wellness states, above and beyond what the data from wearables or demographic tells us. Clearly, social networks are an important piece of the puzzle about our health and wellness. We showed that by including features derived from social networks, accuracy increases significantly and at times using only social network features adds more predictability. Specifically, we find that happiness and positive attitude have the most significant jump when using social network structure features in addition to health behavior and demographic data. This clearly demonstrates that it is the tight coupling of an ego's social and health behaviors that result in improved understanding and predictability of the ego's wellness state.

There are additional insights that might also be gleaned by our study. Consider the correlation among structural variables and health behavior variables (see, e.g. table \ref{spear}). We observe a moderate to strong correlation between clustering coefficient and heart rate, steps, and high activity states which may capture participation in campus sports. These activities provide participating students with ample amounts of physical activity and tightly knit social groups, factors which have been previously shown to be associated with mental health \cite{rosenquist2011social,morgan2013exercise}. Further, it seems that it is easier for social network structure to capture the activity states when a person is either in an inactive state or at least fairly active, than if the person is lightly active. It could be indicative of the relationship between activity and gregariousness or extraversion of an individual. Also, as lightly active minutes include walking, the location of dorms, classes and other necessary destinations involved in a students daily routine may contribute significant noise to this level of activity. A future research direction is looking at more granular data and time windows to understand the immediacy of communication patterns with respect to activity states. 

\bibliography{main}


\section*{Acknowledgements}
The research reported in this paper was supported by the National Heart, Lung, and Blood Institute (NHLBI) of the National Institutes of Health under award number R01HL117757. The content is solely the responsibility of the authors and does not necessarily represent the official views of the National Institutes of Health.

\section*{Author contributions statement}

SL. and NVC. conceived of the research. SL, PRG and NVC developed the modeling framework. SL implemented the models and conducted the experiments. SL, LF, PRG, and NVC performed the analyses.  SL, LF, PRG, and NVC wrote the manuscript.

\section*{Additional information}
\textbf{Supplementary information:} 
All the materials needed to evaluate the conclusions in the paper are available in the paper and/or the Supplementary Materials. Additional information related to this paper may be requested from the authors.

\noindent\textbf{Data Availability:}
All the data related to the paper can be requested from the authors.

\noindent\textbf{Competing interests:}The authors declare that they have no competing interests.





\end{document}